# Exploring the Theoretical Limits of Efficiency in Multilayer Solar Cells

Jeonggyu Hwang[1, *]

[1]*Department of Semiconductor Engineering,*
*Gachon University, Seongnam-si, Gyeonggi-do, South Korea 13120*

Photovoltaic materials are recognized for their potential as sustainable energy sources that enable the conversion between light and electrical energy. However, solar cells have been unable to surpass the theoretical limit of 32%, known as the Shockley-Queisser limit, and face challenges in effectively utilizing the broad spectrum of sunlight. To address this issue, extensive research is being conducted on multi-junction solar cells, which employ a layered structure comprising materials with varying bandgaps to more effectively harness the wide spectrum of sunlight. This study calculates the theoretical limit of these multi-junction solar cells and identifies optimal bandgap combinations, exploring new possibilities for photovoltaic devices and suggesting directions for technological advancement. The performance saw a 31% improvement when moving from a single layer to two layers, and a 12% improvement from two layers to three, with the average wavelength situated in the late 800nm range. The wavelength of the bottom layer increased by about 200nm, and that of the second layer by about 150nm. Our findings present new opportunities to surpass the current limitations of solar cell technology, potentially enhancing the economic feasibility and utility of solar energy as a sustainable source.

Photovoltaic materials are recognized as sustainable energy sources that enable the conversion between light and electricity. However, the efficiency of photovoltaic materials is limited by the theoretical limit known as the Shockley-Queisser limit, which caps at 32%[1]. This limitation arises from the problem that a single material cannot effectively utilize a broad spectrum of solar radiation[2].

Against this backdrop, tandem solar cells, which feature a multilayer structure, are proposed as a promising alternative capable of overcoming the Shockley-Queisser limit and enhancing efficiency. Tandem solar cells employ layers of materials with different bandgaps, stacked together, to effectively harness a wide spectrum of sunlight. By enabling each layer to absorb different parts of the solar spectrum, overall efficiency can be increased.

In this study, we focus on the theoretical limits of solar cells with a multilayer structure. This research systematically analyzes the standard irradiance to find the optimal bandgap combination and predict the theoretical maximum efficiency. Through this approach, we aim to explore new possibilities beyond the efficiency limits of photovoltaic devices and suggest directions for the future development of solar cell technology. We strive to contribute to enhancing the economic viability and practicality of solar power as a sustainable energy source.

## Methodology
### Performance Calculation of Solar Cells

By using solar cells with a multilayer structure, it is possible to reduce thermalization loss that occurs from the energy difference when photons are absorbed, compared to solar cells that use materials with a narrow bandgap. Additionally, using a material with a wide bandgap as the absorption layer can decrease transmission loss caused by the penetration of low-energy photons[3]. Therefore, in tandem solar cells, materials with relatively higher bandgap energies such as perovskites are used on the top[4], while materials with relatively lower bandgap energies like silicon and CIGS (Copper Indium Gallium Selenide) are mainly used on the bottom.

Therefore, this study aims to determine the optimal bandgap based on data analysis of standard solar irradiance. It does not take into account reflection losses, losses due to internal resistance of the device, and surface recombination losses. After calculating the bandgaps for both single-junction and multi-junction solar cells, the theoretical maximum efficiency was determined.

### Data Collection and Preprocessing

The ASTEM (The American Society for Testing and Materials) G-173 spectrum is a dataset provided by the National Renewable Energy Laboratory in the United States, which measures the spectral irradiance of sunlight under specified surface orientations and atmospheric conditions. This dataset is used as an important reference for evaluating the performance of photovoltaic materials. The spectrum is based on measurements under Air Mass 1.5 conditions, where Air Mass 1.5 represents the amount of solar radiation passing through the atmosphere when the sun is at an angle of approximately 48.19 degrees from the horizon. This corresponds to conditions in the middle of the day at an angle approximately equivalent to the average latitude of the continental United States. Detailed spectral irradiance for light wavelengths in Fig. 1.

---

* h5638880@gachon.ac.kr

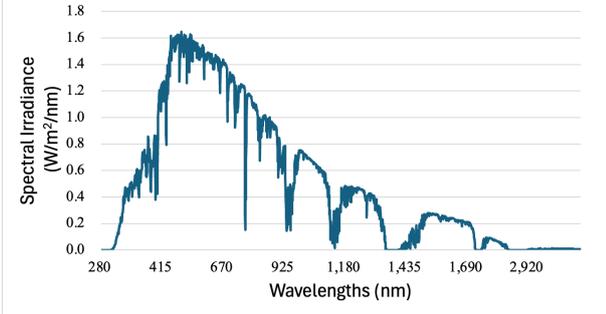

Figure 1. Detailed spectral irradiance for light wavelengths.

This dataset provides values for the wavelengths of light and their spectral irradiance, so we need to convert the wavelengths of light into units of electron volts. Electron volts is represented as

$$E = \frac{hc}{\lambda}\frac{1}{e} \tag{1}$$

where $h$ is Planck's constant; $c$ is the speed of light in vacuum; $\lambda$ is the wavelength in meters; $e$ is the charge of an electron.

## Calculation Algorithm

This algorithm was developed by modifying brute force search methods studied in cryptography to design the optimal bandgap for multilayer structures by evaluating all possible combinations for each layer using a given dataset to find the optimal combination. In this process, an array called the k-array, which represents the starting points of each layer, is used to explore all possible combinations. Each element of this array must be greater than or equal to the previous element.

In the algorithm, the $k_{array}$ starts at $k_{min}$ and can increase up to $k_{max}$, with the length of the array determined by the number of layers. This algorithm does not simply calculate all possible combinations but sequentially generates and evaluates combinations. During this process, as specific combinations are evaluated, some elements of the k-array are repeatedly increased.

## Limitation of Calculation Frequency

Analyzing the rate of increase in the number of calculations, it can be observed that the required number of calculations increases exponentially as the number of layers in the simulation increases. In a single-layer simulation, approximately 2 million calculations are performed, which increases to about 1.3 billion calculations when moving to a two-layer simulation. Furthermore, for a three-layer simulation, the number of calculations dramatically rises to about 670 trillion. According to this trend, the number of calculations required for a four-layer simulation is expected to surge to over 240 trillion. This marks a significantly high figure compared to the previous layers, indicating the complexity of the calculations and the magnitude of the resources required. Therefore, due to the limitations of computational resources, this study was only able to discuss the results up to a three-layer simulation.

The calculations were performed using Google Cloud Platform. The calculations were executed on an E2-micro instance (with 2 vCPUs, 1 core, and 1GB memory). Each calculation took 1 minute for single-layer simulation, 10,100 minutes (approximately 7 days) for two-layer simulation, and 10,200 minutes (approximately 7 days) for three-layer simulation. The computation time did not increase significantly compared to the number of calculations for two-layer and three-layer simulations. This is presumably because the use of Python modules such as Numba allowed storing and recalling past calculation results, which prevented the computation speed from increasing proportionally to the number of calculations.

## Results
### Bandgap Modeling and Simulation

Detailed calculation results in Figure 2. and the maximum values for each layer are shown in Table. 1. In the single-layer simulation, the highest value was achieved at 895nm (approx. 1.3852eV), with a value of 0.811941W/m².

In the two-layer simulation, the highest values were found at the 2nd layer 1109nm (approx. 1.1179eV) and the 1st layer 612nm (approx. 2.0258eV). With these parameters, the output increased by about 31%

(a) 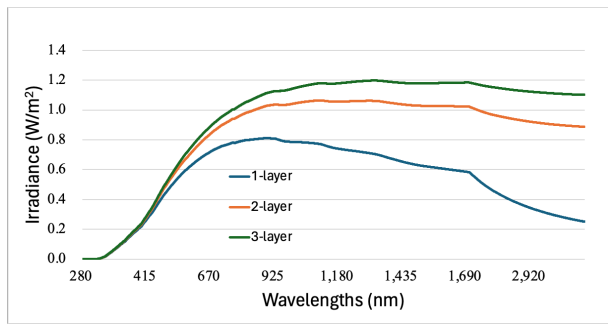

(b) 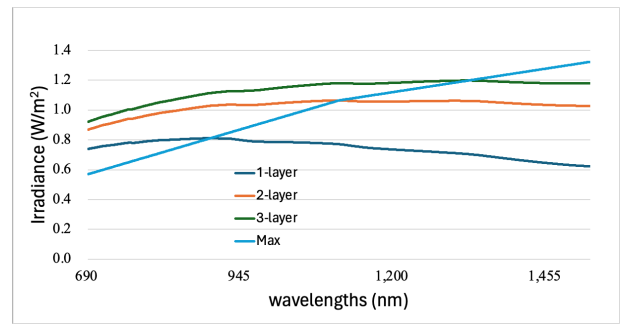

Figure 2. Maximum performance by combination according to the lowest layer wavelength.
(a) Full range (b) 690-1530nm range

| 1-layer | 2-layer | 3-layer | Maximum values |
|---------|---------|---------|----------------|
| 895nm   |         |         | 0.811941W/m$^2$ |
| 1109nm  | 612nm   |         | 1.064789W/m$^2$ |
| 1325nm  | 759nm   | 513nm   | 1.198258W/m$^2$ |

Table 1. Maximum values for each layer.

compared to the single-layer simulation, reaching 1.064789W/m$^2$.

In the three-layer simulation, the highest values were found at the 3rd layer 1325nm (approx. 0.9357eV), the 2nd layer 759nm (approx. 1.6335eV), and the 1st layer 513nm (approx. 2.4168eV). With these parameters, the output increased by about 12% compared to the two-layer simulation, reaching 1.198258W/m$^2$.

**Result Analysis**

As the number of layers increased, the results increased. This proves that a multi-layer structure can absorb more photons effectively. The results have the following correlations.

Firstly, it was found that the average wavelength is located in the late 800nm range. In the case of a single layer, it is 895nm, for two layers it is 860.5nm, and for three layers, it is 865.6nm on average. Thus, it can be evaluated that there is a high probability of achieving the maximum result when the average wavelength is located in the late 800nm range.

Secondly, it was observed that the wavelength of the bottommost layer increases by approximately 200nm. The wavelength of the bottommost layer increased by 214nm when moving from a single layer to two layers, and it increased by 216nm when moving from two layers to three layers. Therefore, it can be evaluated that there is a high probability of achieving the maximum result when the bottommost layer increases by about 200nm as the layers increase.

Lastly, it was confirmed that the wavelength of the second layer increases by about 150nm. When increasing from two layers to three layers, the wavelength of the second layer increased by 147nm. Thus, it can be assessed that there is a high probability of achieving the maximum result value when the second layer increases by about 150nm as the number of layers increases.

**Discussion**

The results indicate that multilayer structures can effectively increase the efficiency of solar cells by minimizing thermalization and transmission losses. The study confirms the theoretical possibility of surpassing the Shockley-Queisser limit with tandem solar cells. However, practical implementation requires addressing material challenges[5].

Future research should focus on experimental validation of the theoretical models and exploring new materials that can achieve the optimal bandgap combinations identified[6]. For instance, materials such as perovskites have shown promise due to their adjustable bandgaps and high absorption coefficients. Furthermore, integrating advanced manufacturing techniques, such as solution processing and vapor deposition, could facilitate the practical realization of these optimized structures[7].

Another critical aspect to consider is the impact of real-world environmental factors on the performance of multilayer solar cells. Factors such as temperature fluctuations, shading, and spectral variations due to atmospheric conditions can influence the efficiency and longevity of photovoltaic devices[8]. Developing robust models that incorporate these factors will enhance the reliability and applicability of the theoretical predictions.

**Conclusion**

This study presents the wavelengths and corresponding bandgaps that yield the maximum efficiency for solar cells with multi-layer structures. It evaluates the performance of all possible combinations for each layer to minimize thermalization loss and transmission loss, and we have found the optimal combination of bandgaps for each layer. As the number of layers increased, the performance improved by 31% (when transitioning from a single layer to two layers) and by 12% (when transitioning from two layers to three layers). The average wavelength was located in the late 800nm range. The wavelength of the bottom-most layer increased by approximately 200nm, and the wavelength of the second layer increased by about 150nm. This groundbreaking research provides a foundation for approaching the efficiency limits of photovoltaic devices.

The insights gained from this study can inform the design and development of next-generation solar cells that not only surpass the Shockley-Queisser limit but also offer practical solutions for large-scale solar power generation[9].